\documentclass{aa}
\pdfoutput=1 
\usepackage{appendix}
\usepackage{latexsym}
\usepackage{graphicx} 
\usepackage{lscape}
\usepackage{afterpage}
\usepackage{booktabs}
\usepackage{multirow}
\usepackage{arydshln}
\usepackage{ulem}

\usepackage{ulem}
\usepackage[citecolor =blue]{hyperref}
\usepackage{psfrag,color,ulem,gensymb}
\definecolor{deepblue}{rgb}{0,0,0.5}
\definecolor{deepred}{rgb}{0.6,0,0}
\definecolor{deepgreen}{rgb}{0,0.5,0} 
\definecolor{darkgreen}{rgb}{0,0.6,0}
\definecolor{darkspringgreen}{rgb}{0.09, 0.45, 0.27}
\definecolor{debianred}{rgb}{0.84, 0.04, 0.33}

\newcommand{\msun}{M$_{\rm \odot}$~}

\DeclareRobustCommand{\ION}[2]{%
\relax\ifmmode
\ifx\testbx\f@series
{\mathbf{#1\,\mathsc{#2}}}\else
{\mathrm{#1\,\mathsc{#2}}}\fi
\else\textup{#1\,{\mdseries\textsc{#2}}}%
\fi}

\usepackage{amstext}

\begin{document}
\title{AGN feedback can produce metal enrichment on galaxy scales}
\author{M. Villar Mart\'{i}n$^{1}$, C. L\'opez Cob\'a$^2$, S. Cazzoli$^3$, E. P\'erez Montero$^3$, A. Cabrera Lavers$^{4,5}$\thanks{Based on observations made  at  Paranal Observatory under programme ID 0102.B-0107}}
\institute{$^{1}$Centro de Astrobiolog\'{i}a (CAB), CSIC-INTA, Ctra. de  Ajalvir, km 4, 28850 Torrej\'{o}n de Ardoz, Madrid, Spain\\
$^2$Institute of Astronomy and Astrophysics, Academia Sinica, No. 1, Section 4, Roosevelt Road, Taipei 10617, Taiwan\\
$^3$Instituto de Astrof\'\i sica de Andaluc\'\i a, CSIC, Apartado de correos 3004, E-18080 Granada, Spain \\
$^4$GRANTECAN, Cuesta de San Jos\'e s/n, 38712 Bre\~na Baja, La Palma, Spain \\
$^5$Instituto de Astrof\'\i sica de Canarias, V\'\i a  L\'actea s/n, 38200 La Laguna, Tenerife, Spain \\
\email{villarmm@cab.inta-csic.es}}        
\date{}
\abstract 
{Giant ($>$100 kpc) nebulae associated with active galaxies provide rich information about the circumgalactic medium (CGM) around galaxies, its link with the interstellar medium (ISM) of the hosts and the mechanisms involved in their evolution.}
{We have studied the giant nebula associated with the Teacup ($z$=0.085) quasar based on VLT MUSE integral field spectroscopy to investigate whether the well known  giant ($\sim$10 kpc) active galactic nucleus (AGN) induced outflow has an impact on the distribution of heavy elements in and outside the host galaxy.}
{We have mapped the oxygen and nitrogen gas relative abundances (O/H and N/O) in two spatial dimensions across the giant nebula and within the galaxy by means of comparing  emission line ratios with photoionisation model predictions.}
{The widely studied AGN driven outflow responsible for the $\sim$10 kpc ionised bubble is enhancing the gas metal abundance up to $\sim$10 kpc from the AGN. O/H is solar or slightly higher in the bubble edges, in comparison with the   subsolar abundances across the rest of the nebula median  (O/H$\sim$0.63 (O/H)$_ {\rm \odot}$).  }
{AGN feedback can  produce metal enrichment at large extranuclear distances in galaxies ($\ge$10 kpc).}

\keywords{galaxies: evolution --  galaxies: active -- galaxies: abundances -- quasars: individual: Teacup  }

\titlerunning{AGN feedback and metal enrichment }
\authorrunning{Villar-Mart\'\i n et al.}

\maketitle
\nolinenumbers

\section{Introduction}

The circumgalactic medium (CGM) is the bound gas halo surrounding galaxies outside their interstellar medium (ISM) and inside their virial radius, extending out to a few hundred kpc.  The CGM may be the key regulator of the galactic gas supply. Gas flows occurring between the CGM and the interstellar medium (ISM) are thought to shape galaxies and drive their evolution, via feedback, accretion and recycling of gas.  Thus, investigating the metallicity, structure and kinematics of the different gas phases will help us to understand how galaxies gain, eject and recycle the gas during their existence  (see \citealt{Tumlinson2017} for a review).

Because the CGM is very diffuse and therefore almost invisible in emission, its physical properties  remain largely unconstrained. Its  understanding  has come so far mostly  from studies based on absorption lines produced by the CGM around galaxies, imprinted on the spectrum of background objects such as quasars and radio galaxies.  The CGM of a few  non-active nearby galaxies has also been studied in emission (e.g. \citealt{Hayes2016}), but the procedure is extremely challenging. The presence of a powerful active galactic nucleus (AGN) can render the CGM gas observable in emission around galaxies up to many 10s of kpc, well into the CGM.  Giant emission line  nebulae (size $\ga$60 kpc and  sometimes $>$100 kpc) associated with  quasars and radio galaxies  at different redshifts have been studied since the 80's  (e.g.\citealt{Baum1988,McCarthy1990,McCarthy1993,vanOjik1996,Villar2003,Borisova2016,Villar2018,Fossati2021,Balmaverde2022,W2023}). 
The discovery  of giant ($\ge$70 kpc) widely spread reservoirs of molecular gas (e.g. \citealt{Emonts2016,Falkendal2021}) associated with several high $z$ ($z$$\ga$2) radio galaxies reveals  a multiphase CGM that was chemically enriched when the Universe was 3 Gyr old and supplies   gas  from which galaxies grow.

Wide field integral field spectroscopic instruments  on 8-10 meter telescopes, such as the Multi Unit Spectroscopic Explorer (MUSE) on the Very Large Telescope (VLT),  have opened excellent opportunities to detect and study in great detail the elusive material from the warm ($T<$10$^5$ K) ionised CGM around powerful active galaxies at different redshifts thanks to the illumination by the powerful AGN. If it were not for the excitation by the AGN continuum, this gas might not be detected or would only be observable through absorption line studies.  

The object of this work is  the  well known Teacup  radio quiet type 2 quasar at $z=$0.085.  It shows a  $\sim$10 kpc loop of ionised gas resembling a handle of a teacup (hereof its nickname), which was discovered by volunteers of the Galaxy Zoo project (\citealt{Keel2012}) and has been widely studied in the context of AGN feedback and its potential impact on galaxy evolution.  The system has been proposed to be the scenario of a giant outflow generated either by an AGN wind or induced by a 1 kpc radio jet whose effects are noticed up to at least $\sim$10 kpc from the AGN and might be responsible for the bubble-like morphology  (\citealt{Gagne2014,Harrison2015,Keel2015,Ramos2017,Villar2018,Moiseev2023}; Venturi et al. (2023, hereafter \citealt{Venturi2023}). 

\cite{Villar2018} discovered  a $>$100 kpc ionised nebula associated with this object (see also \citealt{Villar2021}, which could be a product of a  merger that occurred 1-2 Gyr ago (\citealt{Keel2015}).  This rich gas reservoir, which extends into the CGM, has been rendered visible due to the activity of the quasar  nucleus.  AGN photoionisation dominates the excitation of the spatially resolved gas emission up to its  boundary  (\citealt{Gagne2014,Villar2018,Venturi2023,Moiseev2023}), except at some locations outside the putative quasar ionisation cones where evidence for shock excitation has been found (\citealt{Venturi2023}).  Stellar photoionisation could also contribute to ionise the gas locally in some tidal features. The large scale  kinematics are strongly reminiscent of rotation (\citealt{Villar2018,Moiseev2023}) and tentative results suggest subsolar nebular  abundances ($\sim$0.5 $Z_{\rm \odot}$, \citealt{Villar2018}). The well known bubble appears to be expanding from the nucleus and out into the nebula. 

We present in this paper a detailed optical spectroscopic study of the Teacup nebula  based on VLT-MUSE archival  data with the main goal of  mapping the gas abundances in two spatial dimensions. The ultimate science goal is   to establish whether the giant outflow has an impact on the distribution of heavy elements from the nucleus on large spatial scales. We will discuss the results in the context of other studies of the CGM  and its role in the evolution of galaxies.

The paper is organised as follows: the data and analysis method are presented in Sect. \ref{data} and \ref{analysis} respectively. The latter includes the description of the methods used to derive the gas chemical abundances and physical properties of the gas. The results are presented in Sect. \ref{results} and discussed in Sect. \ref{discussion}. The main conclusions are summarised in Sect. \ref{conclusions}.

Throughout this paper, we assume flat $\Lambda$CDM cosmology
following \cite{Planck2020}, with H$_0$ = 67.4 km s$^{-1}$ Mpc$^{-1}$, $\Omega_{\rm m}$ = 0.31. This gives
a spatial scale of 1.65  kpc arcsec$^{-1}$ at $z=0.085$.

\section{Data}
\label{data}

The  data were collected for the 0102.B-0107 program (principal investigator, PI: L. Sartori; see \cite{Venturi2023} for details) with the  European Southern Observatory (ESO) Very Large Telescope (VLT) and the Multi Unit Spectroscopic Explorer (MUSE, \citealt{Bacon2010}). This instrument covers a 1$\arcmin\times$1$\arcmin$ field of view (FoV) in the Wide Field Mode (WFM), with a spatial sampling of   0.2$\arcsec$ pix$^{-1}$. The wavelength coverage is  $\sim$4650-9300 \AA\, with a 1.25 \AA\ pix$^{-1}$ spectral sampling  and  a resolving power $R=\lambda$/$\Delta\lambda\sim$1700-3400 ($\Delta V\sim$176-88 km s$^{-1}$).

The observations were performed in  March  2019. The processed archive data cube was used for this study.  The Teacup ionised nebula  fills a large fraction of the MUSE FoV.  Separate sky cubes are not available from these observations. Sky oversubtraction due to contamination of the sky spectrum by object emission was identified at a few spatial locations of very low surface brightness  which appeared as artefacts that mimic  absorption features adjacent to the strongest emission lines (specially [OIII]$\lambda$5007). Because our attempts to improve the sky subtraction did not achieve significant improvements, we finally used the   archive cube. To further evaluate the potential impact of this effect, a comparison with the processed archive datacube from program 0103.B-0071 (PI: C. Harrison) was also performed. Although separate sky cubes were obtained for this program, the resulting sky subtraction was not significantly better.  The comparison was in any case valuable to  characterise the impact of the sky subtraction on the data quality. The artefacts are in general faint in comparison with the emission lines and are shifted in wavelength so that they could be efficiently identified and isolated.  We confirm that imperfections on the sky subtraction do not affect our results and conclusions.  These are all based on the 0102.B-0107 datacube  because of the significantly better seeing (FWHM=0.74$\arcsec$ vs. 1.3$\arcsec$ for the 0103.B-0071 program) and of somewhat better signal-to-noise ratio (SN).  

\section{Analysis}
\label{analysis}

\subsection{Spatially resolved emission line flux measurements}

Our main goal is to map the  electron temperature, $T_{\rm e}$, and the oxygen  abundance O/H in two spatial dimensions to investigate whether the giant outflow has an impact on the distribution of heavy elements across the galaxy and out into the CGM. For the $T_{\rm e}$ determination,  it is essential to measure the flux ratio of a nebular to an auroral emission line, such as the [OIII]$\lambda$5007/$\lambda$4363 
(or [NII]$\lambda$6583/$\lambda$5755). For this, we defined spatial apertures  at different locations through the nebula and extracted the integrated spectrum from each one. 

\begin{figure*}
\centering
\includegraphics[width=1.0\textwidth]{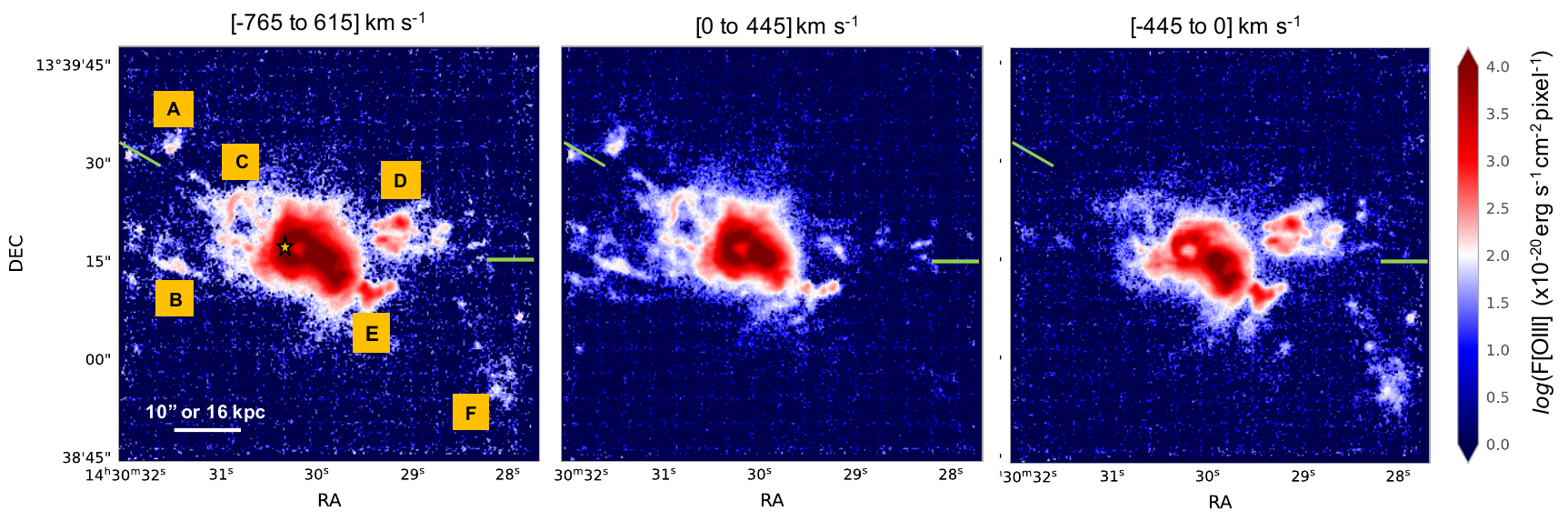}
\caption{[OIII] continuum subtracted images covering different spectral windows that were selected to highlight the diversity of nebular morphological features.  Each image covers a different velocity (i.e. spectral) range relative to the nuclear systemic velocity as indicated on top. The nebular morphology  strongly varies with velocity.  The left panel shows  the total [OIII] flux narrow band image.  The well known $\sim$10 kpc ionised bubble is marked with a tiny yellow star in the left panel. The green lines indicate the position angles of the radio axis to the NE and  to the W (\citealt{Harrison2015}). To guide the reader, the letters `A' to `F'  mark some emission line features that can also be identified in the mask map of  Fig. \ref{kinematicsnebula}. }
\label{imagesnebula}
\centering
\includegraphics[width=1.0\textwidth]{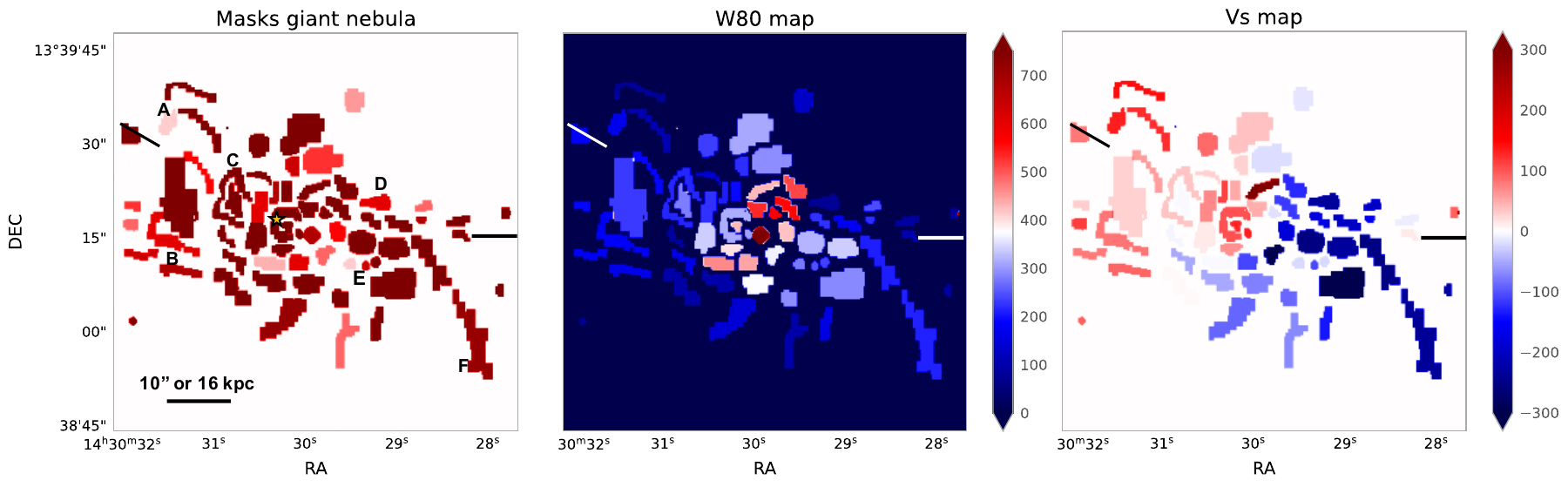}
\caption{Map of the masks used in our analysis (left) and [OIII] kinematic maps (middle and right panels). The colours in the first  map have no particular meaning but help  differentiating the apertures. A 1D spectrum was extracted from each one, so that a single $W_{\rm 80}$ and  $V_{\rm s}$  (middle and left panels) values are associated to each aperture. $V_{\rm s}$ is the velocity shift relative to the narrow core of the nuclear [OIII] line. The  maps cover the total MUSE FoV ($\sim$1$\arcmin\times$1$\arcmin$). $W_{\rm 80}$ and  $V_{\rm s}$   are in km s$^{-1}$. Letters `A' to `F' , yellow star and solid lines have the same meaning as in Fig. \ref{imagesnebula}.}
\label{kinematicsnebula}
\end{figure*}

The apertures  were selected based on the visual inspection of the  [OIII]$\lambda$5007 morphology at different wavelengths  (i.e. velocities) scanned through the line profile. 
These scans reveal striking  morphological  changes with velocity   and are specially useful to identify some  faint structures, such as tidal tails, arcs, knots, filaments, which have similar velocities (Fig. \ref{imagesnebula}).  By restricting the central $\lambda$ and the width, $\Delta\lambda$,  of the narrow band images, the contrast of the lowest surface brightness features is enhanced and the spatial apertures covering them can be defined with greater accuracy to maximise the SN of the integrated spectra. This was useful to detect faint emission lines through the largest possible extension, while preserving at the same time the morphological information on the diversity of nebular structural elements.

The map of the positions of the resulting 64 selected apertures is shown if Fig. \ref{kinematicsnebula} (left panel).  We also show in adjacent panels of the same figure the $W_{\rm 80}$ and $V_{\rm s}$ maps based on [OIII]$\lambda$5007. $W_{\rm 80}$ is  the velocity width that encloses
80\% of the total line flux and  $V_{\rm s}$ is the velocity shift relative to the narrow core of the nuclear [OIII] line, considered here as indicator of the systemic velocity (\citealt{Greene2005}). These maps  can be directly compared with those of \cite{Venturi2023}, that were based on a spaxel by spaxel analysis. The huge dimensions of the nebula both along and perpendicular to the radio axis are apparent. Although some prominent features stand out (knots, filaments, etc), gas emission seems to fill the entire area within the nebular outer boundaries.  The large scale rotation pattern (\citealt{Gagne2014,Villar2018,Moiseev2023}) and the enhanced width, $W_{\rm 80}$, of the lines in the direction perpendicular to the radio axis are clear (\citealt{Venturi2023}).

The 1D spectra extracted from the individual apertures were used to measure the  fluxes of the lines involved in our analysis: [OIII]$\lambda\lambda$4959,5007 and $\lambda$4363, H$\gamma$, H$\beta$, H$\alpha$, [NII]$\lambda\lambda$6548,6583, and $\lambda$5755 and [SII]$\lambda$6716,6731.

For a subset of spectra the stellar continuum was relatively strong compared with the emission line fluxes, that is, the equivalent width of the lines was low, so that it was  first necessary to fit it and subtract it. The effects of stellar absorption  can   affect the Balmer emission lines and need to be accounted for to obtain accurate reddening estimates. It was also necessary in several spectra to reconstruct the H$\gamma$ and  [OIII]$\lambda$4363   baseline and recover both line fluxes. 	The continuum was fitted  for each spectrum with the {\sc pypipe3d} tool \citep[e.g.,][]{Lacerda2022}, that performs a decomposition of the 
observed stellar spectra into multiple simple stellar populations, each of different age and metallicity. As a test, we also subtracted the continuum with 
 the method described in \cite{Cazzoli2022}. The results were consistent. We  also checked the reddening inferred from H$\gamma$/H$\beta$ and H$\alpha$/H$\beta$ in the final, corrected spectra.   Case B  H$\alpha$/H$\beta$=2.87 and  H$\gamma$/H$\beta$=0.47  were assumed. Reddening estimations from both ratios  are in general consistent  within the errors. Changing these values within the range H$\alpha$/H$\beta$=2.76-3.05 and H$\gamma$/H$\beta$=0.45-0.47   (\citealt{Osterbrock1989}) has no significant impact. The  $E_{B-V}$ values implied by H$\alpha$/H$\beta$ were used for extinction correction since they have smaller errors. Using instead the  $E_{B-V}$ implied by H$\gamma$/H$\beta$  has also a negligible impact on the results.

\subsection{Derivation of temperatures and gas chemical abundances}
\label{tempmet}

We used the python code {\sc PyNeb} (\citealt{Luridiana2015}) to estimate  $T_{\rm e}$ of [OIII] for spectra with detected [OIII]$\lambda$4363 using [OIII]$\lambda$5007/$\lambda$4363, and,  when possible, of [NII] using [NII]$\lambda$6583/$\lambda$5755. This code was also used to infer the electron density $n_{\rm e}$ using  [SII]$\lambda\lambda$6717,6731. We calculated the corresponding errors by applying to the calculations a Monte Carlo iteration using the nominal fluxes perturbed with the  observational errors.

For the derivation of  abundances, the direct method (i.e. the derivation of total chemical abundances from their ionic fractions and the measured electron temperature) cannot be satisfactorily applied for the case of the Narrow Line Regions (NLR) in Active Galactic Nuclei (AGN) (e.g. \citealt{Dors2015}). Instead, we used the code  {\sc HII-CHI-mistry} (hereafter {\sc HCm}, \citealt{Perez2014}).   The  approach of this code consists of a bayesian-like comparison between certain reddening-corrected ratios of emission line fluxes relative to a recombination H line (H$\beta$ in our case), that are sensitive to total oxygen abundance (O/H), nitrogen-to-oxygen ratio (N/O), and ionisation parameter ($U$), with the predictions from a large grid of photoionisation models.
In particular, we used version 5.3 of {\sc HCm}, which considers models
 calculated under the most usual conditions in the NLR of active galaxies. As a result, the most probable values and their uncertainties for O/H, N/O and $log(U)$ are obtained.

The grid of models is explained in detail in \cite{Perez2019}.   The gas is assumed to be distributed homogeneously with a filling factor of 0.1 and a constant electron density of $n_{\rm e}$=500 cm$^{-3}$. The densities across the Teacup nebula are however significantly lower ($n_{\rm e}\la$150 cm$^{-3}$, see Table \ref{models} and  \citealt{Venturi2023}).
 Since collisional de-excitation effects are not relevant for the line ratios under consideration in this density regime,  the inferred abundances are not affected by the comparatively high density used in the models (\citealt{Perez2019}). 

The Spectral Energy Distribution (SED) was considered to be composed by two components: one representing the Big Blue Bump peaking at 1 Ryd, and the other a power law with spectral index $\alpha_{\rm OX}$=-1.2 representing the non-thermal X-rays radiation (see \citealt{Perez2019} for additional details). The stopping criterion to measure the resulting emergent spectrum is that the proportion of free electrons in the ionised gas is lower than 2\%.  The models consider  a \cite{Mathis1977} grain size distribution and a dust-to-gas mass
ratio of 7.5$\times$10$^{-3}$ (\citealt{Remy2014}).

The grid of models vary the oxygen and nitrogen abundances within 12+log(O/H)=6.9-9.1 in bins of 0.1 dex and  log(N/O)=-2.0-0.0 in bins of 0.125 dex, respectively. The rest of chemical abundances are scaled in the models with respect to oxygen following the solar proportions, 12+log(O/H)=8.69$\pm$0.04, given by (\citealt{Asplund2021}). In accordance with these authors, we also assumed  12+log(N/H)=7.83$\pm$0.07 for the Sun.

The original grid of models in \cite{Perez2019} covers  values of $log(U)$ from -4.0 to -0.5 in bins of 0.25 dex, although we  only assumed $log(U)>$-2.5 to break the degeneracy with $U$ for the ratios of high- to low-excitation lines in the optical range predicted by the models. This restriction does not imply any significant change in the derivation of the chemical abundances  (see \citealt{Perez2019} for a better clarification). The code also provides uncertainties to the derived quantities calculated as the quadratical addition of the standard deviation of the obtained bayesian distribution with the uncertainty obtained from a Monte Carlo iteration through the given input nominal flux of each line  perturbed with its corresponding observational error.

\begin{figure*}
\centering
\includegraphics[width=1.0\textwidth]{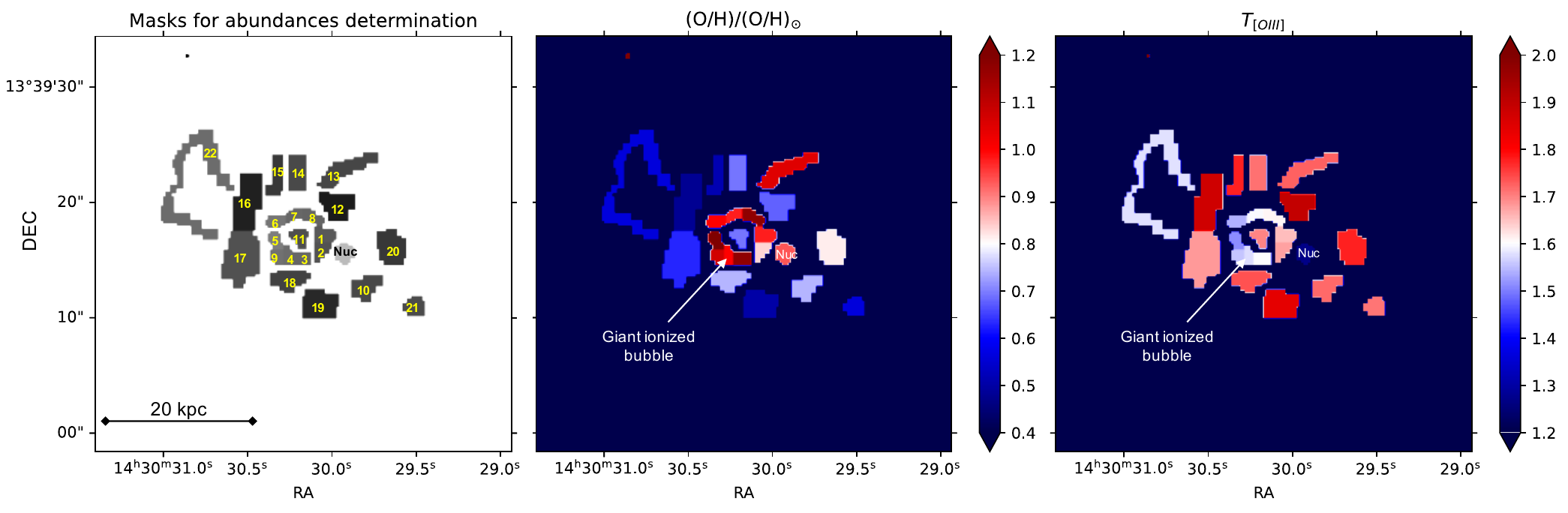}
\caption{Maps of (O/H)/(O/H)$_{\rm \odot}$  ratio and $T_{\rm [OIII]}$. The masks used in this analysis are shown in the left panel. They correspond  to the apertures for which [OIII]$\lambda$4363 is detected. Notice that the  FoV  is smaller than in Fig. \ref{imagesnebula} and \ref{kinematicsnebula}. The exact values of O/H and $T_{\rm [OIII]}$ are in Table \ref{models}. $T_{\rm [OIII]}$ is in units of 10$^4$ K. }
\label{metallicity}
\end{figure*}
\begin{figure}
\centering
\includegraphics[width=0.5\textwidth]{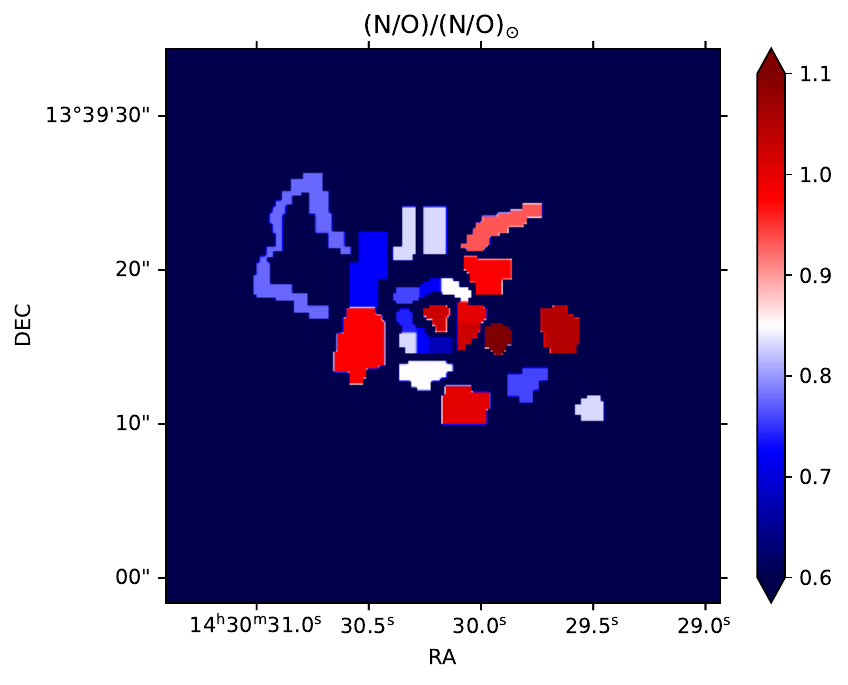}
\caption{Map of (N/O)/(N/O)$_{\rm \odot}$. The FoV is the same as in Fig. \ref{metallicity}. The nuclear value is 1.99, which is outside the colour bar range to enhance the contrast for visualisation purposes. The (N/O)/(N/O)$_{\rm \odot}$  values  are   in Table \ref{models}.}
\label{nitrogen}
\end{figure}

\section{Results}
\label{results}

We show in Table \ref{models}   the O/H and N/O abundances predicted by  photoionisation models  for different apertures, as well as  the   [OIII] and [NII]  electron temperatures inferred with {\sc pyNeb} for the densities estimated with [SII]$\lambda\lambda$6716,6731  (see Sect. \ref{tempmet}).  The observed extinction corrected  line ratios relative to H$\beta$ and measured with the 1dim spectra are also shown. This analysis was performed  for a  set of 23 (including the nucleus) apertures where [OIII]$\lambda$4363 has been detected (see left panel in Fig. \ref{metallicity}).  
The top (lower) part of the table comprises the apertures in which [NII]$\lambda$5755 is detected (undetected). This does not affect the estimated chemical abundances, but simply indicates whether $T_{\rm [NII]}$ could be inferred or not. The  O/H, N/O and $T_{\rm [OIII]}$ information in Table \ref{models}  is also shown as 2dim maps in Fig. \ref{metallicity} and \ref{nitrogen}.

The main result which stands out in Fig. \ref{metallicity} is that the giant ionised bubble shows significantly higher O/H and lower  $T_{\rm [OIII]}$  than the gas at most locations across the giant  nebula. This is confirmed in all apertures along the bubble edge, where $12+log$(O/H) is in the range   8.68$\pm$0.09-8.77$\pm$0.12. As in the nucleus (8.66$\pm$0.11), these abundances are consistent with the  solar value  or  slightly higher.  

For the rest of the nebula (this is, the extended gas outside and beyond the bubble), except in Ap. 13,   the oxygen abundances are  subsolar everywhere (as tentatively found by \citealt{Villar2018}). Considering all apertures across the nebula, the $log$ of the median is 12+$log${\rm (O/H)}=8.49 (or 63\% solar), with values as low as  8.37$\pm$0.08-8.44$\pm$0.13 (48\%-56\% solar)    in Ap. 15, 16, 19, 21, 22.  Interestingly, the gas encircled by the bubble edges (Ap. 11), with  12+$log$(O/H)=8.53$\pm$0.17 (69\% solar), shows an intermediate abundance between the most enriched gas (the bubble edges) and the more metal poor gas in the rest of the  nebula.

It is important to highlight that the abundance predictions are based on the assumption that the gas is photoionised by the AGN everywhere. This is reasonable for most apertures except, possibly, at a few positions  located in the direction perpendicular to the main axis of the nebula, where shocks might be contributing to the excitation of the gas    (\citealt{Venturi2023}). These apertures are labelled Ap. 12, 18, 19 in Fig. \ref{metallicity}, left panel.  In spite of this uncertainty, the fact that (except for Ap. 13) this gas follows  the general behaviour of the O/H and $T_{\rm [OIII]}$  maps suggests that the derived abundances are reliable. 

In general, the $T_{\rm [OIII]}$  map (right panel in Fig. \ref{metallicity}) mimics the behaviour of the  O/H maps in an anticorrelation. The nucleus shows the minimum temperature, $T_4$=$T_{\rm [OIII]}$/10$^4$=1.27$\pm$0.08 K, in comparison with the rest of the gas, which shows $T_4>$1.5 everywhere.  As before, the bubble stands out in this map, with the edges being  significantly colder ($T_4\sim$1.5-1.6 in Ap. 3 to 9) than  the giant nebula, which is very hot with $T_4\sim$1.7-1.9 almost everywhere. \cite{Venturi2023} presented a 2dim $T_{\rm [OIII]}$ map tracing the nucleus and the ionised bubble.  They inferred a narrow range of  significantly lower $T_{\rm [OIII]}\sim$(1.3-1.4)$\sim$10$^4$ K compared to our estimations. The reason for this discrepancy remains unknown.

In regard to the N/O abundance (Fig. \ref{nitrogen} and Table \ref{models}), the most obvious result is the much higher nuclear value ($log$(N/O)=-0.56$\pm$0.08, i.e. roughly twice the solar abundance ratio of -0.86$\pm$0.07), compared with the solar or somewhat below solar N/O everywhere else (-1.03 $\pm$0.09$\le log{\rm (N/O)}\le$-0.84$\pm$0.20), including the bubble edge. There is not such a striking   trend of N/O with  location as found for O/H. The bubble edge shows among the lowest N/O  ratio ($\sim$68-85\% N/O$_{\rm \odot}$), although not unique in comparison with other regions of the giant extended nebula. The bubble edge N/O values are  lower than the gas encircled by it (ap. 11), the nucleus  and the gas in between (ap. 1 and 2).

$T_{\rm [NII]}$  could be measured only for the nucleus  and up to the bubble edge (Table \ref{models}). The temperature difference between these two regions is less pronounced than for $T_{\rm [OIII]}$. $T_{4 ~\rm [NII]}\sim$0.983$\pm$0.066 K in the nucleus and somewhat higher in the bubble ($\sim$1.01-1.12 K).

\section{Discussion}
\label{discussion}

With a maximum  extension  as traced by the MUSE data of $\sim$126 kpc  (this is a lower limit since the gas fills the FoV in this direction),  the Teacup giant nebula traces part of the CGM. Its properties are strongly conditioned by the AGN, but it still provides valuable information about the CGM.  If it was not for the nuclear activity, most (if not all) of this gas would  remain invisible.  

 The giant nebula shows subsolar abundances, with O/H$\sim$(48\%-84\%)$\times$(O/H)$_{\odot}$ almost everywhere, and median  63\%, well below the nuclear, roughly solar abundance. For comparison,  different works  based on absorption line studies have shown that the dense gas in the CGM of $z\la$1 galaxies has a bimodal metallicity distribution function, with an equal number of absorbers in the low-metallicity ($Z\la$0.03$Z_{\odot}$) and high-metallicity ($Z\sim$0.4$Z_{\odot}$) branches (e.g. \citealt{Lehner2013,Wotta2016}). The abundance of the Teacup ionised nebula falls in the latter  group (clearly, this does not discard the presence of lower metallicity gas). The high-metallicity branch has been proposed to trace galactic winds, recycled outflows, and tidally stripped gas (\citealt{Lehner2013}).

Lower CGM metallicities compared with the ISM have been found  for different galaxy types, based  on studies of absorption line systems at $\la$200 kpc from their host galaxies  (e.g. \citealt{Kacprzak2019}).  
These authors found a large range of abundances in the CGM of isolated star forming galaxies, 0.01$\la$ Z/Z$_{\odot}<1$  and an offset  of $log (dZ)$=-1.17$\pm$0.11 between the CGM and ISM, which shows no dependence with stellar mass. The relation has a large scatter of 1$\sigma$=0.72. The offset for the Teacup is smaller for O/H ($ log (dZ) \sim$-0.17 using the median nebular O/H), although still within the scatter. Different processes may be at work in this system, related to the nuclear activity and/or to its merger history.

We have shown that the $\sim$10 kpc Teacup bubble to the East of the nucleus, that is known to be driven by an AGN wind or the small nuclear radio jet, is associated with  obvious changes in the gas abundance.  The bubble edge shows significantly higher O/H  (solar or slightly solar, similar to the nucleus) in comparison with the subsolar O/H across the rest of the nebula.  
Most likely as a consequence, the bubble edge is also significantly less hot ($T_{4}\sim$1.5-1.6) than the rest of the nebula (1.7$\la T_{4}\la$1.9). 
Therefore, the outflow appears to be causing a change in the  gas metal content from the nucleus  up to $\sim$10 kpc.

This  mechanism may also explain the nuclear deficit of O/H. For the Teacup values of log($M_{\rm *}/M_{\odot}$)=11.15$\pm$0.05   and  star forming rate $SFR\sim$10 \msun yr$^{-1}$ (\citealt{Jarvis2020,Ramos2022}),  12+log(O/H)$\sim$8.85 is expected, according to the mass-metallicity-SFR ($M_{\rm*}$-Z-SFR) relation by \cite{Perez2013} or $\sim$8.8 for the extrapolation of the  $M_{\rm *}$-Z-SFR relation by \cite{Andrews2013} to high  $M_{\rm *}$.  Seyfert 2 galaxies with similar  $M_{\rm *}$   tend to show O/H  consistent with these predictions (\citealt{PerezDiaz2021}). The Teacup nucleus, on the contrary, shows lower than expected O/H (8.66$\pm$0.05), 
while the nuclear N/O  (-0.56$\pm$0.08, Table \ref{models}) is consistent with that expected for its $M_{\rm *}$ ($log$(N/O)=-0.54, \citealt{Perez2013,Andrews2013}). 

The dilution produced by inward flows of low-metallicity gas (for instance, from the outskirts of the two merging galaxies) could explain the  nuclear  O/H and N/O   (\citealt{Edmunds1990,Rupke2010}). The outflow is also  an interesting possibility.  This scenario is  supported by the similar O/H of the bubble edge and the nuclear gas, which is moreover enhanced in comparison with the rest of the nebula. On one hand,  radiative outflows can couple more efficiently with metals via resonance line scattering  (e.g. \citealt{Pauldrach1994,Arav1994,Higginbottom2024}; see also \citealt{Edmunds1990}). On the other, galaxies show metallicity gradients such that the inner regions have higher abundances (\citealt{Searle1971,Kewley2019,Maiolino2019}). Based on this, a gradient is expected to exist within  the $\sim$4 kpc diameter nuclear aperture used in the Teacup analysis. The outflow has been generated in the inner regions ($\la$1 kpc) close to the AGN (\citealt{Harrison2015,Ramos2017,Venturi2023}), where the gas is expected to be more metal rich. If it drags gas out to large distances, as proposed in this scenario, the global metallicity of the residual gas within the  nuclear aperture would be lower as a consequence.
  
An implication of this scenario is  that the outflow has been capable of ejecting gas  from the galaxy center and dragged it up to $\sim$10 kpc. If the outflow expanded without displacing significant amounts of gas to large distances, the bubble abundance would be similar to the rest of the nebula. The implications are important. It supports that  metal-enriched galactic outflows (driven by an AGN in this particular case) shape the mass-metallicity relationship, by  removing metals from galaxy potential wells and ejecting them to large distances, possibly out into the CGM  (\citealt{Tremonti2004,Peeples2014,Chisholm2018,Tortora2022}). 
 
The  behaviour of N/O remains to be explained. An outflow could preserve  N/O (\citealt{Edmunds1990} (specific modeling would be valuable for AGN generated winds). On the contrary,  this ratio is depleted in the bubble edge  in comparison with the nucleus and its gradient shows no obvious spatial correlation with the bubble, but just a tentative trend to show among the lowest  N/O. This  is not necessarily a discrepancy.  Given the complexity of the N/O behaviour in terms of  secondary and primary stellar production processes, and the fact that the ejected gas would mix    with gas across the nebula with a  non-uniform N/O distribution, it is difficult to predict how this ratio would behave as the bubble expands and mixes with the pre-existing reservoir.

 An alternative scenario to gas ejection from the centre is that local chemical enrichment  has been produced by young stars. This  is supported by the detection of blue-coloured continuum emission co-spatial with the bubble edge  due to  a population of stars that are younger ($\la$100-150 Myr) than in the rest of the galaxy ($\ga$0.5-1 Gyr, \citealt{Venturi2023}). According to these authors,  widespread star formation has been  triggered at the edge of the bubble due to the compressing action of the jet and outflow (positive feedback).  The time scale could be long enough to enrich the local gas with O, but not with secondary N  (\citealt{Molla2006,Kumari2018}). This would explain the enhanced O/H in comparison with the rest of the nebula, while  N/O is not clearly different, and being at the same time significantly lower than in the nucleus. 

Yet another possible scenario to explain the abundance values in the bubble relates to the depletion by dust of metals from the gas-phase ISM. If shocks destroyed dust as the bubble  expands (e.g. \citealt{Dopita2018}),  metals could be released to the gas. Since oxygen is more sensitive to depletion than nitrogen (\citealt{Jones2019}), this could explain  the higher O/H of the bubble in comparison with the rest of the nebula, while having a tentatively lower N/O.

Whether due to nuclear gas ejection, to local star formation or dust destruction, in all three scenarios the AGN induced outflow  is responsible for the metal enrichment of the gas at distances as large as $\sim$10 kpc.  The implications are different. In the ejection scenario, the behaviour of O/H provides observational evidence of how AGN induced outflows can deprive the central regions of galaxies from metals and transfer them up to very large distances, possibly out of the galaxy and into the CGM. In the second scenario,  the behaviour of O/H provides evidence of how AGN induced outflows can produce local metal enrichment (this is also the case in the third, the dust depletion scenario) at very large distances from the nucleus, with  a delay between the quasar onset and the induced metal enrichment  of $\sim$100-150 Myr.

\section{Conclusions}
\label{conclusions}

The giant ($\ga$126 kpc) nebula associated with the Teacup QSO2 at $z=$0.085 traces part of its CGM. Its properties are strongly influenced by the nuclear activity up to the outer detected emission line regions, where it still provides valuable information about the CGM surrounding the quasar host galaxy.  If it was not for the nuclear activity, most (if not all) of this gas would  remain invisible.  This study is an example of the great potential of studying  giant nebulae  to investigate in emission  the  CGM around active galaxies at all redshifts.

The widely studied AGN driven outflow responsible for the well known ionised bubble is enhancing the gas metal content (O/H) up to $\sim$10 kpc from the AGN. The giant nebula shows subsolar metallicity almost everywhere, except the bubble, which has about solar or slightly super-solar metallicity.

This could be a consequence of the ejection of metal rich gas from the nucleus. In such scenario, the Teacup provides observational evidence for how  AGN feedback  can deprive the central regions of galaxies from gas and displace metals out to very large distances, possibly out of the galaxy.  It supports that metal-enriched AGN outflows can shape the mass-metallicity relationship of galaxies. 
Alternatively, the O/H enrichment could have been produced locally by the young stellar population    formed in the bubble edge, possibly formed as a consequence of positive feedback $\sim$100-150 Myr (\cite{Venturi2023}). A third possibility is the release of oxygen to the gas  phase as a consequence of dust destruction in the bubble by shocks triggered by the expanding outflow.

In any of the scenarios considered,  the nuclear activity is the ultimate mechanism responsible for  the metal enrichment of the gas at large extranuclear distances ($\sim$10 kpc).

\begin{acknowledgements} 
We are grateful to  Luc Binette, Bjorn Emonts and Bruno Rodr\'\i guez for  useful feedback. We also thank the referee for the careful revision of the manuscript and valuable suggestions. MVM and ACL research has been funded by grant Nr. PID2021-124665NB-I00  by the Spanish Ministry of Science and Innovation/State Agency of Research MCIN/AEI/ 10.13039/501100011033 and by "ERDF A way of making Europe". SC acknowledges financial support from the Severo Ochoa grant CEX2021-001131-S and the Ministry of  Science, Innovation and Universities (MCIU) under grants PID2019-106027GB-C41. EPM acknowledges financial support  by project Estallidos8 PID2022-136598NB-C32 (Spanish Ministry of Science and Innovation).   The Cosmology calculator by \cite{Wright2006} has been used.

\end{acknowledgements}

\begin{appendix}
\begin{landscape}
\begin{table}
\centering
\tiny
\caption{Photoionisation model predictions} 
\begin{tabular}{clllllllllllllll}
\hline
Aperture &	 E(B-V) &        [OIII]$\lambda$4363   &          [OIII]$\lambda$5007 &          [NII]$\lambda$5755 & [NII]$\lambda$6584 &  [SII]$\lambda$6725  &    [SII]$\frac{\lambda 6716}{\lambda 6731}$       & 12+$log$(O/H)      &    $log$(N/O)     &           $T_{\rm [OIII]}$   & $T_{\rm [NII]}$  & $n_{\rm e}$\\ 
&	&	&	&	&	&	&	&	& & 	K~~~ &	K~~~ &	cm$^{-3}$	\\
\hline
1  &   0.074$\pm$0.012 & 0.144$\pm$0.014 & 6.000$\pm$0.215 & 0.019$\pm$0.001 & 1.437$\pm$0.114 &   1.598$\pm$0.047 & 1.316$\pm$0.056 & 8.68$\pm$0.15 &  -0.86$\pm$0.11 & 16565$\pm$875  & 9508$\pm$312 &  149$\pm$79\\
2 &  0.155$\pm$0.015 & 0.141$\pm$0.012  & 5.802$\pm$0.588 & 0.017$\pm$0.001 & 1.396$\pm$0.044 & 1.507$\pm$0.051 &   1.396$\pm$0.026  &  8.61$\pm$0.13 & -0.85$\pm$0.11 & 16739$\pm$1196  & 9236$\pm$225 &  34$\pm$36 \\
3 &  0.054$\pm$0.013 & 0.135$\pm$0.015 & 6.043$\pm$0.251 &0.020$\pm$0.002 & 1.044$\pm$0.042 &1.467$\pm$0.060 &  1.349$\pm$0.048  &     8.76$\pm$0.11 & -1.03$\pm$0.09 &   16058$\pm$1071  & 11185$\pm$448 & 100$\pm$63\\
4 & 0.034$\pm$0.013 & 0.158$\pm$0.012 & 7.290$\pm$0.251 & 0.018$\pm$0.002 & 1.012$\pm$0.048 & 1.359$\pm$0.071 & 1.382$\pm$0.027 &    8.69$\pm$0.10 & -1.00$\pm$0.08 &   15801$\pm$699  &  10822$\pm$543 & 51$\pm$34\\
5 &  0.063$\pm$0.018 & 0.142$\pm$0.008 & 7.199$\pm$0.322 & 0.016$\pm$0.002 & 0.991$\pm$0.062 & 1.196$\pm$0.056 & 1.343$\pm$0.033 &    8.77$\pm$0.12 & -0.99$\pm$0.08 &   15064$\pm$493 & 10325$\pm$565 & 109$\pm$41\\
6 &  0.046$\pm$0.017 & 0.143$\pm$0.010 & 6.885$\pm$0.319 & 0.019$\pm$0.001 & 1.086$\pm$0.063 & 1.326$\pm$0.078 & 1.321$\pm$0.036 &     8.69$\pm$0.12 & -0.98$\pm$0.11 &   15433$\pm$641 & 10656$\pm$344 &  141$\pm$52 \\
7 & 0.101$\pm$0.014 & 0.162$\pm$0.009 & 7.342$\pm$0.323 & 0.019$\pm$0.001 & 1.070$\pm$0.067 & 1.419$\pm$0.059 & 1.341$\pm$0.036   &  8.68$\pm$0.09 & -1.00$\pm$0.08 &  15938$\pm$501 & 10743$\pm$353 & 114$\pm$50 \\
8 & 0.087$\pm$0.021 & 0.128$\pm$0.014 & 5.665$\pm$0.326 & 0.020$\pm$0.001 &1.291$\pm$0.112 & 1.589$\pm$0.088 &1.369$\pm$0.046&   8.76$\pm$0.15 & -0.93$\pm$0.11 &  16181$\pm$1037 & 10134$\pm$391 & 74$\pm$60\\
9 & 0.031$\pm$0.013 & 0.145$\pm$0.006 & 6.900$\pm$0.316 &0.020$\pm$0.002 &1.195$\pm$0.069 & 1.423$\pm$0.072 & 1.394$\pm$0.021  &   8.72$\pm$0.10 & -0.94$\pm$0.10 &  15453$\pm$442 & 10495$\pm$516 & 40$\pm$33 \\
10 & 0.105$\pm$0.016 & 0.115$\pm$0.019 &4.557$\pm$0.208 & 0.020$\pm$0.002 & 1.169$\pm$0.079 & 1.632$\pm$0.074 &   1.411$\pm$0.016 &   8.56$\pm$0.20 & -0.98$\pm$0.11 &  16888$\pm$1484 & 10580$\pm$530 & $\la$35 \\
 Nucleus & 0.186$\pm$0.033 & 0.086$\pm$0.014 & 6.550$\pm$0.310 & 0.021$\pm$0.003 & 1.497$\pm$0.157 & 0.920$\pm$0.059 & 1.242$\pm$0.138 &   8.66$\pm$0.11 & -0.56$\pm$0.08 &   12772$\pm$785  & 9825$\pm$658  & 301$\pm$270\\
\hline
11 & 0.000$\pm$0.046 & 0.146$\pm$0.021 & 5.941$\pm$0.387&  N/A &1.271$\pm$0.183 & 1.455$\pm$0.082 & 1.335$\pm$0.017 & 8.53$\pm$0.17 & -0.85$\pm$0.15 & 16899$\pm$1339 & N/A & 115$\pm$22\\ 
12 & 0.132$\pm$0.015  & 0.113$\pm$0.014  & 3.570$\pm$0.170 & N/A & 1.350$\pm$0.061 &1.865$\pm$0.085  &  1.450$\pm$0.050 &  8.52$\pm$0.14 &-0.87$\pm$0.11 & 19540$\pm$1235 & N/A & 16$\pm$34 \\
13 & 0.000$\pm$0.038 & 0.083$\pm$0.023 & 3.197$\pm$0.325 &   N/A&  1.378$\pm$0.166 & 1.973$\pm$0.228  &  1.417$\pm$0.054 &  8.71$\pm$0.24 & -0.89$\pm$0.14 & 17766$\pm$2661 & N/A & 34$\pm$47 \\
14 & 0.031$\pm$0.025 & 0.175$\pm$0.035 & 6.901$\pm$0.531 &  N/A & 1.244$\pm$0.104 & 1.684$\pm$0.126  &  1.330$\pm$0.029 &  8.53$\pm$0.15 & -0.94$\pm$0.11 & 17197$\pm$2124  & N/A & 129$\pm$42 \\
15 & 0.077$\pm$0.041 & 0.198$\pm$0.024 & 7.222$\pm$0.808 &    N/A& 0.914$\pm$0.115 & 1.480$\pm$0.183  &  1.316$\pm$0.025 &  8.40$\pm$0.12 & -0.94$\pm$0.13 & 17989$\pm$1652 & N/A & 145$\pm$34 \\
16 & 0.023$\pm$0.015 & 0.271$\pm$0.040 & 8.844$\pm$0.387  &  N/A & 0.647$\pm$0.072 & 1.382$\pm$0.319  & N/A & 8.37$\pm$0.08 & -1.00$\pm$0.11 & 18829$\pm$1648 & N/A  & N/A \\ 
17 & 0.029$\pm$0.020 & 0.222$\pm$0.017 & 8.941$\pm$0.404   &  N/A& 1.123$\pm$0.102 &1.339$\pm$0.146  & 1.429$\pm$0.061 & 8.49$\pm$0.05 & -0.87$\pm$0.12 & 16910$\pm$754& N/A & 33$\pm$48\\
18 &  0.121$\pm$0.012 & 0.102$\pm$0.023 & 3.917$\pm$0.210 &  N/A & 1.097$\pm$0.058& 1.602$\pm$0.191 &   1.466$\pm$0.030 &  8.56$\pm$0.24 & -0.93$\pm$0.13 & 17299$\pm$2055 & N/A  & $\la$75\\
19 &  0.027$\pm$0.044 &0.090$\pm$0.028 & 2.626$\pm$0.481 & N/A & 1.245$\pm$0.283 & 1.125$\pm$0.378 &  	1.386$\pm$0.039 & 8.39$\pm$0.30 & -0.86$\pm$0.28 &  18925$\pm$3439 & N/A & 48$\pm$45 \\
20 & 0.077$\pm$0.031 & 0.087$\pm$0.016 &  3.075$\pm$0.420 & N/A & 1.494$\pm$0.282 & 1.888$\pm$0.284 & 1.396$\pm$0.039 & 8.60$\pm$0.25 & -0.84$\pm$0.20  &  18398$\pm$2835 & N/A & 41$\pm$45 \\ 
21 & 0.126$\pm$0.059 & 0.201$\pm$0.033 & 8.234$\pm$1.289   &  N/A& 0.716$\pm$0.126 & 0.947$\pm$0.164 &  1.372$\pm$0.028 & 8.44$\pm$0.13 & -0.94$\pm$0.16 & 16972$\pm$1929 & N/A &  70$\pm$38\\
22 & 0.000$\pm$0.007 & 0.154$\pm$0.012 & 7.099$\pm$0.143 &   N/A & 0.776$\pm$0.038 & 1.164$\pm$0.338  &  N/A & 8.44$\pm$0.18 & -0.97$\pm$0.17 & 15798$\pm$608 & N/A & N/A \\ \hline
\end{tabular}
\tablefoot{Model predictions of the O/H and N/O relative abundances, the electron temperatures, $T_{\rm [OIII]}$ and  $T_{\rm [NII]}$ and electron densities $n_{\rm e}$ inferred from the  [SII]$\frac{\lambda 6716}{\lambda 6731}$ ratio. The observed, extinction corrected line fluxes used in the models are quoted relative to H$\beta$.  E(B-V) values were inferred from H$\alpha$/H$\beta$. The spatial location of the apertures can be seen in Fig. \ref{metallicity}. Apertures 3 to 9 cover the edge of the ionised bubble. The top (lower) part of the table comprises the apertures in which [NII]$\lambda$5755 is detected (undetected). N/A in the [SII]$\frac{\lambda 6716}{\lambda 6731}$ column means that the doublet ratio could not be measured accurately due to residual sky  artefacts. The solar  abundances are  12+$log$(O/H)=8.69$\pm$0.04 and          $log$(N/O)=-0.86$\pm$0.07 for the Sun (\citealt{Asplund2021}).} 
\label{models}
\end{table}
\end{landscape}
\end{appendix}

\end{document}